\documentstyle[12pt,epsfig]{article}

\newlength{\dinwidth}
\newlength{\dinmargin}
\newcommand{\ba}{\begin{array}}
\newcommand{\ea}{\end{array}}
\newcommand{\ben}{\begin{equation}}
\newcommand{\een}{\end{equation}}
\newcommand{\bea}{\begin{eqnarray}}
\newcommand{\eea}{\end{eqnarray}}
\newcommand{\nn}{\nonumber}
\newcommand{\gsim}{\mathrel{\mathop{\kern 0pt \rlap
  {\raise.2ex\hbox{$>$}}} \lower.9ex\hbox{\kern-.190em $\sim$}}}

\def\bz{{\bar z}}
\def\bt{{\bar t}}
\def\bu{{\bar u}}
\def\bw{{\bar w}}

\def\btau{{\bar \tau}}
\def\bphi{{\bar \phi}}
\def\baeta{{\bar \eta}}
\def\bpsi{{\bar \psi}}
\def\tit{{\tilde t}}
\def\tr{{\tilde r}}

\def\bchi{{\bar \chi}}
\def\cO{{\cal O}}

\def\pt{t^\prime}
\def\pz{z^\prime}
\def\pr{r^\prime}
\def\bt{{\bar t}}
\def\bz{{\bar z}}

\def\tSigma{{\tilde \Sigma}}
\def\ttau{{\tilde \tau}}
\def\tphi{{\tilde \phi}}
\def\hr{{\hat r}}
\def\teta{{\tilde \eta}}
\def\ttheta{{\tilde \theta}}

\begin{document}
\thispagestyle{empty}
\addtocounter{page}{-1}
\begin{flushright}
YITP-01-29 \\
TIFR-TH/01-14\\
Alberta Thy-08-01\\
\end{flushright}
\vspace*{1cm}
\centerline{\large \large\bf Unruh Radiation, Holography and Boundary Cosmology}
\vspace*{1.2cm} 
\centerline{\bf Sumit R. Das${}^{a,b}$ {\rm and} Andrei Zelnikov${}^{a,c,d}$}
\vspace*{0.8cm}
\centerline{\it Yukawa Institute for Theoretical Physics}
\vspace*{0.2cm}
\centerline{\it Kyoto University, Kyoto 606, JAPAN ${}^a$}
\vspace*{0.6cm}
\centerline{\it Tata Institute of Fundamental Research}
\vspace*{0.2cm}
\centerline{\it Homi Bhabha Road, Mumbai 400 005 INDIA ${}^b$
\footnote{Permanent Address}}
\vspace*{0.6cm}
\centerline{\it Theoretical Physics Institute, Department of Physics}
\vspace*{0.2cm}
\centerline{\it University of Alberta, Edmonton AB, CANADA ${}^c$}
\vspace*{0.6cm}
\centerline{\it P.N. Lebedev Physics Institute,}
\vspace*{0.2cm}
\centerline{\it Leninsky pr. 53, Moscow 117924, RUSSIA ${}^d$}
\vspace*{1cm}
\centerline{\tt das@theory.tifr.res.in \hskip1cm zelnikov@phys.ualberta.ca}
\vspace*{1.2cm}
\centerline{\bf Abstract}
\vspace*{0.3cm}

A uniformly acclerated observer in anti-deSitter spacetime is known
to detect thermal radiation when the acceleration exceeds a critical
value. We investigate the holographic interpretation of this
phenomenon. For uniformly accelerated trajectories {\it transverse} to
the boundary of the $AdS$ space, the hologram is a blob which expands
{\it along} the boundary. Observers on the boundary co-moving with the
hologram become observers in cosmological spacetimes.  For
supercritical accelerations one gets a Milne universe when the
holographic screen is the boundary in Poincar\'e coordinates, while
for the boundary in hyperspherical coordinates one gets deSitter
spacetimes.  The presence or absence of thermality is then interpreted
in terms of specific classes of observers in these cosmologies.

\vspace*{1.1cm}

\baselineskip=18pt
\newpage

\section{Introduction}

The concrete realization of the ideas of holography \cite{holography}
in terms of the $AdS-CFT$ correspondence \cite{adscft} has provided a
new insight into spacetime physics. According to this correspondence,
all states of string theory or M-theory in an asymptotically $AdS_m
\times S^n$ background can be described as states 
in a theory which is defined on the boundary of $AdS$ and which does
not involve gravity.  In particular for $m=3,5$ the boundary theory is
a conventional large-N gauge theory. The vacuum of the boundary theory
corresponds to a background which is purely $AdS$ while gauge
invariant states are described as states of string theory defined in
this background. More interestingly, a {\it thermal} state of the
gauge theory is dual to a $AdS$ black hole at the same temperature
\cite{wittenbh}. This is consistent with the success of microscopic
understanding of entropy and Hawking radiation of near-extremal black
holes in string theory
\cite{blackholes}.

The phenomenon of Hawking radiation from Schwarzschild black holes is
deeply related to Unruh radiation \cite{unruh} detected by a uniformly
accelerated observer in flat space, essentially because the
near-horizon geometry of such black holes is Rindler space. A
uniformly accelerated observer in $AdS$ spacetime would also perceive
the $AdS$-invariant vacuum to be a thermal bath, provided the
acceleration exceeds a critical value \cite{deser}, 
a fact which can be proved for any field theory \cite{jacobson}
and can be
easily understood in terms of the trajectories in 
the global embedding spacetime. 
In this paper we attempt a holographic understanding of
such acceleration radiation.

One of the motivations behind this work is to get some holographic
insight into bulk physics as perceived by various classes of
observers. The AdS/CFT correspondence has given a rather satisfactory
description of the bulk physics for a certain class of observers,
e.g. for observers who are stationary outside a black hole. To
understand black hole complementarity, however, we need to compare
this with the description of bulk physics from the viewpoint of other
observers, like those freely falling into a black hole.  Some of these
interesting problems are quite general and do not depend on presence
of a black hole.  In this paper our analysis is focused on observer
dependent properties of quantum fields in $AdS$ spacetime and their
holographic interpretation on the boundary.  We hope that it will provide some
insight into the holographic description of physics for an interesting
class of accelerated observers.

Consider a geodesic in the bulk of $AdS$ which is entirely transverse
to the boundary. As expected from the IR/UV relationship, the ``image''
of this in the boundary theory is a blob \cite{bulkboundary}, or
rather a ``bubble'' \cite{danielsson} which expands in the
longitudinal directions {\it along} the boundary. The size of the
image at any given time is related to the distance of the object from
the boundary. When we have a {\it uniformly} accelerated trajectory
in the bulk we, of course, have a similar physics.  The nature of the
expanding image depends on whether or not the acceleration exceeds a
critical bound.  We will find that observers according to whom the
holographic image is time-independent are like observers in
cosmological spacetimes. When the accleration exceeds the critical
value, such observers preceive the vacuum as a thermal state, while
for subcritical or critical acclerations they dont.

The nature of the cosmology we obtain depends on the particular
coordinates in $AdS$ used to define a boundary. We will need to work
not really on the boundary, but at a small but finite distance away
from the boundary. This implies that the holographic theory has an UV
cutoff. We will nevertheless continue to refer to the holographic
screen as the ``boundary''. For example, on the Poincar\'e boundary,
which is flat, images of observers with a supercritical acceleration
in the bulk, look like observers who use the conformal time in an
expanding Milne universe. 
According to them the image of the
accelerating bulk object is time independent.  These boundary
observers do not cover the complete Poincar\'e boundary and perceive
the invariant vacuum as a thermal state \cite{birel}. The picture
looks different in the case of observers with critical or subcritical
accelerations. These map on to boundary observers whose coordinate
systems cover the entire Poincar\'e boundary.  On the other hand, the
boundary defined in a coordinate system in $AdS_{d+1}$ which is
similar to the standard spherical coordinate system on a sphere and
may be obtained from it by analytical continuation ( we call these
``hyperspherical coordinates'') is $d$ dimensional {\em deSitter}
spacetime.  Now subcritical acclerations correspond to the choice of
global time in $dS_d$. Critical acclerations correspond to choice of
the conformal time in the extension of the steady state universe which
covers the entire $dS_d$ boundary. Supercritical accelerations
correspond to the time in the $dS_d$ coordinate system in which the
metric is time-independent. For well known reasons the last class of
observers detect a thermal bath \cite{gibbhaw, birel}.

The appearance of a cosmological interpretation of Unruh radiation in
the bulk is intruiging. This is similar to the microscopic
interpretation of the deSitter entropy in two dimensions in
\cite{msh}.  Here one has a physical brane in the bulk theory which is
also the holographic screen. Its location is determined dynamically by
the brane tension. Consequently, the holographic theory has dynamical
gravity \cite{gubser,msh} as in brane world models \cite{randall}.
In hyperspherical coordinates, the entropy of the deSitter spacetime
on the screen is understood as entaglement entropy as in theories of
induced gravity \cite{fiola}. From our point of view, this happens
because stationary observers on the screen are bulk accelerated
observers with supercritical acceleration which are moving {\em along}
the screen. In our work, we have a {\em fixed} holographic screen and
we look at objects in the bulk which are away from the screen with
{\em no motion along it}. Nevertheless, examining their holograms lead
to similar connections with cosmological spacetimes - not only for
screens placed near the boundary in hyperspherical coordinates but for
other coordinates as well. This circle of ideas is also somewhat
related to the observation of \cite{deser} that thermal properties of
spacetimes with genuine gravity, like black holes or deSitter spaces
can be in fact derived from the thermal properties of acclerated
observers in global embeddings spacetimes which can be flat. This
latter observation, however, does not use any holographic
correspondence.

In Section 2 we give the definitions of the various coordinate systems
in $AdS_{d+1}$ which are used in the paper.  In Section 3 we derive an
equation which describes the motion of a general uniformly accelerated
object in $AdS$ space and find all solutions to obtain three classes.
In Section 4 we compute the one point functions in
the various boundary theories for $AdS_3$ which provide holograms of
such acclerated objects when they couple to some massless scalar field
in the bulk. In Section 5 we give the boundary interpretation of
presence or absence of thermal behavior as perceived by bulk observers
moving along those trajectories. The various cosmolgies appear
here. In Section 6 we discuss generalizations to higher dimensions and
argue that the connection with boundary cosmologies is
general. Section 7 contains conclusions and comments. Appendix I
contains details of the various coordinate systems. Appendix II
contains the details of the calculation of the formulae used in
Section 4.

\section{Coordinate Systems in $AdS_{d+1}$}

We will consider four classes of coordinate systems in
$AdS_{d+1}$. These describe different ways of choosing $AdS_{d+1}$ as a
hyperbolic surface in $d+2$ dimensional flat space with two timelike
coordinates. The details of the embeddings are given in the Appendix I.

\subsection{Global Coordinates}

The global coordinate system has a time $\tau$ which has a finite
range $0  <  \tau < 2\pi$. However very often one considers a covering
space with an infinite range by attaching identical copies. The
spatial coordinates include a radial variable $\nu, 0  <  \nu <
\pi/2$, and $(d-1)$ other angles $\phi, \theta_i , i=1,\cdots d-2$
with ranges  $0  <  \phi, \theta_i < \pi$ for $i=2,\cdots (d-3)$ and 
$0  <  \theta_{d-2} < 2\pi$. The metric is then given by 
\ben
ds^2 = \sec^2\nu [ - d\tau^2 + d\nu^2 + \sin^2\nu (d\phi^2 + \sin^2 \phi 
d\Omega_{d-2}^2)].
\label{eq:five}
\een
The boundary is at $\nu = \pi/2$ in these global coordinates. Surfaces
of constant $\nu = \nu_B$ have
spatial sections which are $S^{d-1}$. When $\nu_B$ is close to, but
less than ${\pi \over 2}$ the 
induced metric is 
\ben
ds_{gIn}^2 = \sec^2 \nu_B[ -d\tau^2 + \sin^2\nu (d\phi^2 + \sin^2 \phi 
d\Omega_{d-2}^2)].
\label{eq:metrica}
\een
The metric of the boundary theory is obtained by removing the overall
factor of $\sec^2 \nu_B$ from (\ref{eq:metrica}).

\subsection{Poincar\'e Coordinates}

The time in Poincar\'e coordinates has infinite range $-\infty < t <
\infty$ and the spatial coordinates include radial variables $r,z$
with a range $0 < r,z < \infty$ and $d-2$ angles on a $S^{d-2}$. These
are denoted by $\theta_i, i = 1 ,\cdots d-2$ and have the usual ranges 
$0  <  \theta_i < \pi$ for $i=2,\cdots (d-3)$ and 
$0  <  \theta_{d-2} < 2\pi$. These angles $\theta_i$ are common with
the corresponding angles in the global coordinate system. 
The metric is now
\ben
ds^2 = {1\over z^2}[-dt^2 + dz^2 + dr^2 + r^2 d\Omega^2_{d-2}].
\label{eq:eight}
\een
The Poincar\'e system does not cover the entire $AdS$ spacetime. 
For a set of static observers $~\{z,r,\theta_i\}={\mathrm const}~$
there is a horizon at $z=\infty$. 

The Poincar\'e boundary is at $z=0$. Constant $z = z_B$ surfaces are
flat, with induced metric
\ben
ds_{PIn}^2 ={1\over z_B^2}[-dt^2 + dr^2 + r^2 d\Omega^2_{d-2}].
\label{eq:eighta}
\een
The boundary metric is obtained by removing the overall factor
${1\over z_B^2}$.
 
\subsection{``BTZ'' Coordinates}

The third coordinate system we will use will be called ``BTZ''
coordinates, since this is a natural extension of the BTZ coordinates for
$d=2$. The time $\eta$ has infinite range, $-\infty  <  \eta <
\infty$. There is a radial 
variable $\rho$ with $ 1 < \rho <\infty$ and
another noncompact coordinate $\psi$ with $0 < \psi < \infty$  and in
addition the same $d-2$ angles $\theta_i$ as in the global and
Poincar\'e systems.
The metric is now
\ben
ds^2 = -(\rho^2-1)d\eta^2 + {d\rho^2 \over (\rho^2-1)} + \rho^2[d\psi^2  +
\sinh^2\psi d\Omega_{d-2}^2].
\label{eq:eleven}
\een
These coordinates also do not cover the entire $AdS$ spacetime 
and static observers see horizons at $\rho^2 = 1$.

The boundary is at $\rho = \infty$. Finite $\rho =
\rho_B$  surfaces have spatial sections which are
hyperbolic spaces in $(d-1)$ dimensions. For large $\rho_B$  the
induced metric is
\ben
ds^2_{BTZIn} = \rho_B^2[-d\eta^2 + d\psi^2  +
\sinh^2\psi d\Omega_{d-2}^2].
\label{eq:twelve}
\een
The boundary metric is (\ref{eq:twelve}) divided by the factor
$\rho_B^2$. 

\subsection{Hyperspherical coordinates}

These coordinates are obtained from analytic continuations of standard
polar coordinates on a $S^{d+1}$. They include a time which is in fact
identical to the time $\eta$ in BTZ coordinates, a radial variable
$\mu$ with $0 < \mu < \infty$, an angle $\theta$ with range $0 < \theta
< \pi$ and the standard set of $d-2$ angles $\theta_i$ which are
identical to the above. The metric is given by
\ben
ds^2 = -\sinh^2 \mu~\sin^2\theta d\eta^2 + d\mu^2 + \sinh^2 \mu~d\theta^2 +
\sinh^2 \mu \cos^2\theta d\Omega^2_{d-2}.
\label{eq:cone}
\een
Like BTZ coordinates, these  do not cover the entire $AdS$
spacetime. Once again, static observers see a horizon at $\mu = 0$.

The boundary is at $\mu = \infty$. Any finite $\mu = \mu_B$ surface is a
$d$-dimensional deSitter  space with the metric
\ben
ds_{HIn}^2 = \sinh^2 \mu_B (-\sin^2\theta d\eta^2 + d\theta^2 + \cos^2
\theta d\Omega^2_{d-2}),
\label{eq:ctwo}
\een
with a constant Ricci scalar ${6 / \sinh^2 \mu_B}$. Once again the
boundary metric is obtained by removing the overall factor. The Ricci
scalar of the boundary metric is thus $6$.

\section{Accelerated Trajectories}

Consider a trajectory specified by the following equations in Poincar\'e
coordinates 
\ben
z=z(\lambda)~~~~~t=t(\lambda)~~~~~~r=0,
\label{eq:thirteen}
\een
where $\lambda$ denotes the proper time. The components of the
velocity may be then written as
\bea
u^t (\lambda) & = & {dt \over d\lambda} = z(\lambda) \cosh \gamma
(\lambda), \nn \\
u^z (\lambda) & = & {dz \over d\lambda} = z(\lambda) \sinh \gamma
(\lambda),
\label{eq:fourteen}
\eea
with all other components equal to zero. 
We have written $u^t,u^z$ in a way
which automatically satisfies the requirement $g_{\alpha\beta}u^\alpha
u^\beta = -1$.
The nonzero components of an acceleration vector are
\bea
a^t & = & {du^t \over d\lambda} - {2\over z} u^t u^z, \nn \\
a^z & = & {du^z \over d\lambda} - {1\over z} [(u^t)^2 + (u^z)^2].
\label{eq:sixteen}
\eea

\subsection{Uniform Accceleration}

Now we derive a differential equation for $\gamma(\lambda)$ such
that the proper acceleration is constant along the trajectory. This
means that
\ben
g_{\alpha\beta}a^\alpha a^\beta = {1\over z^2}[(a^z)^2 - (a^t)^2]=a^2.
\label{eq:seventeen}
\een
Together with the relation $g_{\alpha\beta}u^\alpha a^\beta = 0$ this
implies that we can write
\ben
a^t = a~z(\lambda) ~\sin\gamma (\lambda),~~~~~~a^z = a~z(\lambda)
~\cos\gamma (\lambda). 
\label{eq:eighteen}
\een
Substituting this in any of the equations in (\ref{eq:sixteen}) we
easily get the following differential equation for $\gamma (\lambda)$
\ben
{d\gamma \over d\lambda} = \cosh \gamma + a .
\label{eq:nineteen}
\een
This equation can be easily solved.

\begin{enumerate}

\item{} In the first solution
the right hand side of (\ref{eq:nineteen}) vanishes,
\ben
\cosh \gamma + a = 0 .
\label{eq:twenty}
\een
which is possible only if $a^2 \geq 1$. $\gamma$ is thus a constant along
the trajectory.
It is trivial to solve for $z(\lambda)$ and $t(\lambda)$. For $a^2 >
1$ we have 
\bea
z(\lambda) & = & A~e^{\pm {\sqrt{a^2-1}}~\lambda}, \nn\\
t(\lambda) & = & \pm {a \over {\sqrt{a^2-1}}} A~e^{\pm
{\sqrt{a^2-1}}~\lambda},
\label{eq:twoone}
\eea
The $\pm$ sign corresponds to the direction of $t$. We can choose this
to be lower sign to get for the trajectory
\ben
t = {a \over {\sqrt{a^2-1}}}~z.
\label{eq:twotwo}
\een
Different values of the integration constant $A$ lead to trajectories
which are related by 
simultaneous rescalings of $z$ and $t$, which is an isometry of the
spacetime . The final
equation (\ref{eq:twotwo}), of course, does not depend on this choice.

When $a = -1$ we have $\gamma = 0$ along the trajectory. It follows from
(\ref{eq:fourteen}) that these trajectories are simply $z ={\mathrm const}$.
 
\item{} The second solution has nonvanishing ${d\gamma \over
d\lambda}$. Using (\ref{eq:nineteen}) and the definitions of $u^\alpha
(\lambda)$ we can now easily solve for $z$ and $t$ as functions of
$\gamma$ 
\bea
z & = & A (\cosh \gamma + a), \nn \\
t & = & A \sinh \gamma .
\label{eq:twothree}
\eea
so that we have the trajectory
\ben
(z-aA)^2 - t^2 = A^2.
\label{eq:twofour}
\een
The solution for
$\gamma (\lambda)$ now depends on whether $a^2 >1,~a^2 = 1$ or $ a^2 < 1$.
For $a^2 > 1$ we have
\ben
\cosh \gamma = {a \cosh [{\sqrt{a^2-1}}~(\lambda -\lambda_0)]-1
\over a - \cosh [{\sqrt{a^2-1}}~(\lambda -\lambda_0)]}~~~~~(a^2 > 1),
\label{eq:twofive}
\een
When $a^2 = 1$ we get
\ben
\coth (\gamma /2) = - (\lambda - \lambda_0)~~~~~(a^2=1),
\label{eq:twosix}
\een
while for $a^2 < 1$ we have
\ben
\cosh \gamma = {a \cos [{\sqrt{1-a^2}}~(\lambda -\lambda_0)]-1
\over a - \cos [{\sqrt{1-a^2}}~(\lambda -\lambda_0)]}~~~~~(a^2 < 1).
\label{eq:twoseven}
\een

\end{enumerate}

For $a^2 \geq 1$ the trajectories of the second type can be in fact
related to those of the first type by isometries of the metric. For
$a^2 > 1$ we first transform to coordinates $(t',z',r',\theta_i)$ defined as
\bea
& U = {\pt \over \pz}, & V = {1\over 2\pz}[1+((\pr)^2+(\pz)^2-(\pt)^2)],
\nn \\ 
& & W_1  =  {1\over 2\pz}[1-(\pr)^2+(\pz)^2-(\pt)^2)], \nn \\ 
& W_2 =  {\pr \over \pz} \cos \theta_1, & 
\cdots 
W_d  =  {\pr\over \pz}\sin\theta_1 \sin\theta_2 \cdots \sin \theta_{d-2}.
\label{eq:twoeight}
\eea
This is clearly an isometry since this involves just interchange of
$U$ and $V$ in (\ref{eq:six}). Defining
\ben
a = {\rho_0 \over  {\sqrt{\rho_0^2-1}}},
\label{eq:twonine}
\een
the trajectory (\ref{eq:twotwo}) is simply
\ben
t = \rho_0 z,~~r=0 ~~~~~~{\rm i.e.}~~~V = \rho_0~~W_n = 0~~(n >1).
\label{eq:thirty}
\een
Then in terms of the primed coordinates this becomes
\ben
(\pz - \rho_0)^2 - (\pt)^2 = \rho_0^2 -1~~~~~\pr = 0.
\label{eq:threeone}
\een
We then make a simultaneous rescaling of $\pz$ and $\pt$
\ben
\bz = {A \over {\sqrt{\rho_0^2-1}}} \pz,~~~~~~~
\bt = {A \over {\sqrt{\rho_0^2-1}}} \pt,
\label{eq:threetwo}
\een
together with a similar rescling of $r$, which is also an isometry. 
The trajectory now takes the form
\ben
(\bz-aA)^2 - \bt^2 = A^2 ,\nn
\een
which is identical to (\ref{eq:twofour}).

\begin{figure}
\centerline{
\epsfig{file=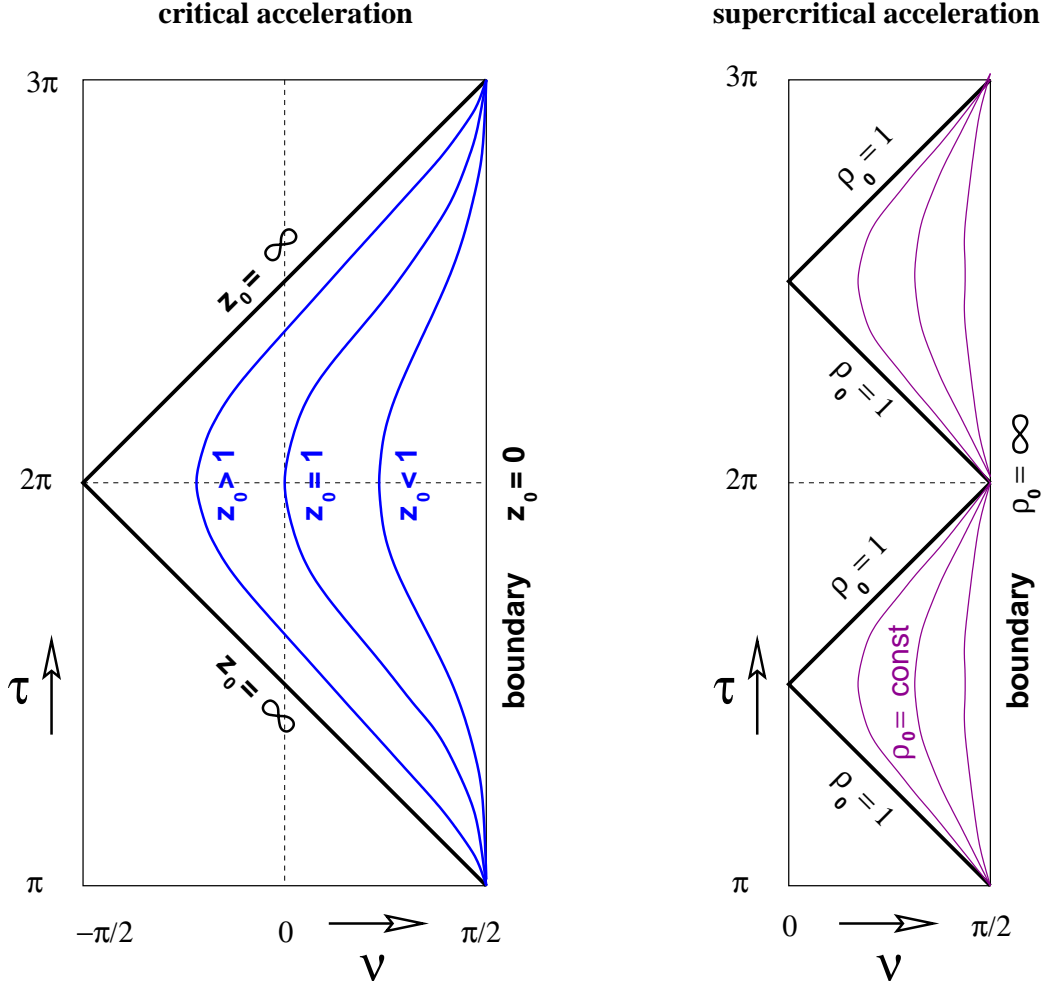,width=14 cm}}
\caption[Fig1]{Left picture: The trajectories corresponding to the critical
accleleration $a=-1$. 
Right picture: The trajectories with supercritical acceleration
$a^2>1$.}
\label{PoincareBTZ}
\end{figure}

Similarly for $a = -1$ the trajectory is given by
\ben
{1\over z_0} = U + W_1,~~~~~~~~r=0  \nn
\een
which becomes
\ben
(\pz + {1\over z_0})^2 - (\pt + 1)^2 = ({1\over z_0})^2. \nn
\een
After performing a shift $\pt \rightarrow \pt + 1$ and a simultaneous
rescaling of $\pt$ and $\pz$ the trajectory becomes identical to
(\ref{eq:twofour}) with $a=-1$ (see Fig.(\ref{PoincareBTZ})).

We therefore need to consider only three classes of trajectories
\bea
(z + \xi_0)^2 -t^2 & = & (1 + \xi_0^2)~~~~~~~~
(a^2 = {\xi_0^2 \over \xi_0^2 + 1} < 1),
\label{eq:trajthree}\\
z & = & z_0~~~~~~~~~~~~~~~~~(a = -1),
\label{eq:trajtwo} \\
t & = & \rho_0 ~z ~~~~~~~~~~(a^2 = {\rho_0^2 \over \rho_0^2 -1} > 1),
\label{eq:trajone} 
\eea
where the first trajectory (\ref{eq:trajthree}) was obtained from
(\ref{eq:twofour}) by choosing $A$ suitably.

Using the definitions of various coordinate systems in the previous
section, it is now easy to see that (\ref{eq:trajone}) corresponds to a
constant value of $\rho = \rho_0$ with $\psi = 0$ 
in the BTZ coordinate system (see Fig.(\ref{PoincareBTZ})), or
equivalently $\mu = \mu_0 = \cosh^{-1} \rho_0$ with $\theta = \pi/2$ 
in the hyperspherical
coordinate system. Similarly,
(\ref{eq:trajtwo}) corresponds to a constant value of $z=z_0$ in the
Poincar\'e coordinate system and (\ref{eq:trajthree}) corresponds to a
constant value of $\nu = \nu_0$ in the global coordinate system
where $\tan \nu_0 = \xi_0$.

We have performed the analysis of accelerated trajectories using the
Poincar\'e coordinate system. However the analysis can be redone in
any other coordinate system. Even though the Poincar\'e, BTZ and
hyperspherical coordinate systems do not cover the entire $AdS$ spacetime
the basic differential equation (\ref{eq:nineteen}) can be easily
transformed to the global coordinate system and solved. The final
result is, of course, the same and we have the three basic classes of
trajectories as listed above. For example, in global coordinates these
trajectories become
\bea
& \nu = \nu_0&~~~~(a^2 <1), \nn \\
& \sec \nu \cos \tau \pm \tan \nu = {1\over z_0}&~~~~(a = -1),\nn \\
& \sec \nu \sin \tau = \rho_0 &~~~~(a^2 > 1).
\label{eq:trajf}
\eea
All these trajectories have $\sin \phi = 0$. Starting from $\phi = 0$
The $\pm$ sign in the second equation refers to the sides of a section
with $\sin \phi = 0$.
Since this trajectory can pass from the side with $\phi = 0$ to the
side which has $\phi
= \pi$, it is conveninent to imagine that the sign of $\nu$ is
reversed in this latter side. This explains why we consider the range
$-{\pi /2} < \nu < \pi/2$ in
Figure (\ref{PoincareBTZ}). 

\subsection{Unruh radiation}

In flat space, any unformly accelerated observer perceives the
Minkowski vacuum as a thermal state. In $AdS$ space, however, among
the three classes of unform acceleration, only observers which follow
the trajectory (\ref{eq:trajone}) or those obtained from this by
isometries will perceive the $AdS$ vacuum as a thermal state
\cite{deser}. A nice way to see this is to consider the corresponding
trajectories in the $d+2$ dimensional space $(U,V,W_i)$ in which
$AdS_{d+1}$ is embedded as a hyperboloid. Only those trajectories with
$a^2 > 1$ have a real acceleration in this embedding spacetime, and
the corresponding temperature can be read off from the usual Unruh
formula \cite{deser}. Another way to see this is to consider the
horizons perceived by the class of accelerated observers and calculate
the surface gravity. This calculation can be done entirely in global
coordinates. However, as in the case of spacetimes with bifurcate
Killing horizons (see \cite{jacobson}), it is
easier done in appropriate coordinates where the time is chosen to be
along the Killing vector which becomes null at the horizon. The
answer is valid for arbitrary interacting fields
and is, of course, coordinate invariant.

Thus for $a^2=1$ trajectories the horizon is
simply at $z = \infty$ (in Poincar\'e coordinates), 
where the Killing vector
$k = z_0{\partial \over \partial t}$ becomes null. 
We assume the Killing vector to be normalized 
$k^\epsilon k_\epsilon=-1~|_{z=z_0}$
at the location of the observer.
The surface gravity for the horizon reads
\ben
\kappa^2 = -{1\over 2}\nabla^\alpha k^\beta \nabla_\alpha k_\beta
|_{z=\infty} = {z_0^2\over z^2} |_{z=\infty} = 0.
\label{eq:threethree}
\een
This implies that the Unruh temperature vanishes.

For the $a^2 > 1$ trajectories are specified by constant $\rho=\rho_0$ 
in the
BTZ coordinates and the horizon is at $\rho = 1$. Here the Killing
vector $k ={1\over\sqrt{\rho_0^2-1}}{\partial \over \partial \eta}$ 
becomes null and the
surface gravity is
\ben
\kappa^2 = -{1\over 2}\nabla^\alpha k^\beta \nabla_\alpha
k_\beta|_{\rho=1}
= {\rho^2\over \rho_0^2-1 } |_{\rho = 1} = {1\over \rho_0^2-1 }
\hskip 1cm 
k^\epsilon k_\epsilon=-1~|_{\rho=\rho_0}.
\label{eq:threefour}
\een
The local temperature measured by an observer at $\rho = \rho_o$
then reads
\ben
T_{local} = {1\over 2 \pi {\sqrt{\rho_0^2 -1}}}.
\label{eq:threesix}
\een
This implies that it is just red-shifted Unruh temperature 
\ben
T_U = {1\over 2\pi},
\label{eq:threefive}
\een
measured at infinity.

For $a^2 < 1$ the trajectories have constant $\nu$ in global
coordinates and there is no horizon and, hence, no temperature.

\section{Holograms of the trajectories in $AdS_3$}

The above trajectories in the bulk have holograms on the boundary. The
profile of the hologram may be obtained as follows. Imagine that we
have a source of some supergravity field which is moving along some
trajectory in the bulk, specified by $Y^\alpha (\lambda)$. The one
point function of the operator of the boundary theory which is dual to
this field evaluated in a state corresponding to the motion of the
source is then obtained by evaluating the value of this field on the
boundary and then integrating over the world-line of the source
\cite{bulkboundary,danielsson}.

For simplicity we will consider a source which is coupled to a
minimally coupled {\em massless} scalar field in the $AdS$
spacetime. We therefore need to calculate the scalar Green's function
between a point on the trajectory $Y^\alpha (\lambda)$ and a point in
the bulk $X^\alpha$ and integrate this over the trajectory. 
Finally we have to take the point $X$ to the boundary.
In this section we will
perform this computation for the case of $AdS_3$.  Generalization to
other $AdS_{d+1}$ and to sources of other supergravity fields should
be quite similar in spirit.

The hologram would of course depend on the specific boundary which is
chosen. We will choose three different boundaries, (i) the boundary
defined in Poincar\'e coordinates defined by $z=0$ , (ii) the boundary
in global coordinates defined by $\nu = {\pi \over 2}$ and (iii) the
boundary defined in hyperspherical coordinates at $\mu = \infty$. 
The one point
functions would be then results in the corresponding boundary theory.
In each case we will consider the holographic screen a little away
from the boundary, i.e. for small but finite $z$, for $\nu$ close
to, but less than ${\pi \over 2}$ and for $\mu$ large but finite. 
As is well known from the IR/UV
correspondence, this corresponds to a boundary theory with an
ultraviolet cutoff.

In $AdS_3$ the scalar Green's function for a massless scalar field is
given by \cite{danielsson}
\bea
G(X,Y) & = & -{1\over 2\pi}\cot (v)~~~~~-1 < \cos v < 1 \nn, \\
& = & 0 \hskip 3.3cm {\rm otherwise},
\label{eq:hone}
\eea
where $v$ is the timelike geodesic distance between the point $X$ and
$Y$.

In global coordinate system defined in (\ref{eq:three}) the geodesic
distance is given by
\ben
\cos (v) = \sec \nu~\sec\nu '~\cos (\tau - \tau ')-\tan \nu~\tan
\nu '~\cos (\phi - \phi '),
\label{eq:htwo}
\een
where we have labelled $X = (\tau,\nu,\phi)$ and $Y = (\tau ',\nu
',\phi ')$.

In Poincar\'e coordinate system defined in (\ref{eq:six}) one has
\ben
\cos (v) = 1 + {(r-r')^2 + (z-z')^2 - (t-t')^2 \over 2 z z'},
\label{eq:hthree}
\een
where $X = (t,z,r)$ and $Y = (t',z',r')$.

In the BTZ coordinate system defined in (\ref{eq:nine}) one
has
\ben
\cos (v) = \rho \rho' \cosh (\psi - \psi') - {\sqrt{(\rho^2-1)(\rho'^2
-1)}}~\cosh (\eta - \eta'),
\label{eq:hfour}
\een
where $X = (\eta, \rho, \psi)$ and $Y = (\eta ', \rho ', \psi ')$.

Finally in the hyperspherical coordinate system, with
$X=(\eta,\mu,\theta)$ and $Y = (\eta', \mu', \theta')$ 
\ben 
\cos (v) =
\cosh \mu\cosh \mu' - \sinh \mu\sinh \mu'[\sin \theta \sin \theta'
\cos (\eta - \eta') - \cos\theta \cos \theta'].
\label{eq:cthree}
\een

The value of the scalar field at the location $X^\alpha$ in the bulk 
due to a
source moving along $Y^\alpha = Y^\alpha (\lambda)$ is then given by 
\cite{danielsson}
\ben
\Phi (X) = \int d\lambda G(X, Y(\lambda)),
\label{eq:hfive}
\een
where we substitute for the specific form of the trajectory $Y^\alpha
(\lambda)$ and $\lambda$ denotes the proper time. 
The answer is of course independent of the coordinate
system used to do the calculation. However the one point function for
a specific boundary theory 
\ben
<\cO> \sim \Phi (X) |_{boundary},
\label{eq:hsix}
\een
{\em depends on the coordinates used to
define the boundary}. The transformation between different coordinate
systems become conformal transformations when restricted to the
boundary. The one point function for different boundary theories can
then be related by transforming $<\cO>$ according to the standard
rules determined by its conformal weight.

Strictly speaking, one has to divide the right side of (\ref{eq:hsix})
by the radial part of the {\em normalizable} wave function evaluated
on the boundary \cite{adscft, bulkboundary, dghosh}.  The latter is a
cutoff factor which is consistent with the correct conformal weight of
$\cO$. However, in the following we will retain this cutoff factor in
the formulae for reasons which will become clear later.

The strategy to calculate $\Phi (X)$ is to first use a coordinate
system which is best suited for the particular kind of
acceleration. Since subcritical accelerations are described in global
coordinates by $\nu = \nu_0$ we use global coordinates for this
computation. Similarly for the critical and supercritical
accelerations we use the Poincar\'e and BTZ coordinates respectively.
Once this is done we can transform to any coordinate system we
like. This latter system is chosen according to which boundary we are
considering. We then take the limit of $\Phi (X)$ for the appropriate
boundary to extract the one point function.

In the following we outline the derivation of boundary one point
functions for Poincar\'e, global and hyperspherical
boundaries. Details of the calculation of the field are given in
Appendix II.

\subsection{Poincar\'e Boundary}

The Poincar\'e boundary is at $z = z_B \rightarrow 0$ and we will need
$\Phi(X)$ in the Poincar\'e coordinate system.

\subsubsection{Subcritical Accelerations, $a^2 < 1$}

First consider trajectories given by (\ref{eq:trajthree}) which have
accelerations $a^2 < 1$. 
Since these correspond to constant values of
$\nu = \nu_0$ it is most conveninent to write down the Green's
function and the field $\Phi (X)$ in global coordinates. The result is
in the Appendix II, equation (\ref{eq:hten}). We then
transform to Poincar\'e coordinates. The result is
\ben
\Phi (X) = {2z \sec\nu_0 \over 2\pi} \int_0^\pi dv~{\cos v \over 
[\sec^2\nu_0
(4t^2 + (1 + s)^2)-(2z\cos v + \tan \nu_0 (1-
s))^2]^{1/2}},
\label{eq:heleven}
\een
where
\ben
s = r^2+z^2 - t^2.
\een
The one point function is then given by
\ben
<\cO>^{a^2<1}_{Poincare}  =  \Phi (X) |_{z = z_B \rightarrow 0}.
\label{eq:htwelve}
\een
The leading answer vanishes and one has to evaluate the
first subleading (in $z_B$) contribution. The
answer is
\ben
<\cO>^{a^2<1}_{Poincare}  =  z_B^2 {\sin \nu_0 \over \cos^2 \nu_0}
{(1-(r^2-t^2)) \over [4t^2 + (1+(r^2-t^2))^2 + 4 r^2 \tan^2\nu_0]^{3/2}}
\label{eq:hthirteen}
\een
Note that there is an overall factor $z_B^2$ in the expression, and $z_B
\rightarrow 0$. From the UV/IR connection $z_B$ appears as a (position
space) UV cutoff in the boundary theory, as is evident from the form
of the metric restricted to the boundary. The power of $z_B$ in
(\ref{eq:hthirteen}) reflects the fact that $\cO$, being dual to a
massless scalar field in the bulk, is a $(1,1)$
operator on the boundary.

\subsubsection{Critical Acceleration}

For the trajectory (\ref{eq:trajtwo}) with critical acceleration one
has $z' = z_0$ in the Poincar\'e system. Therefore we do the
calculation in this coordinate system. The result is equation
(\ref{eq:htwoone}) in Appendix II which we rewrite here
\ben
\Phi (X)  =  {z \over 2\pi}\int_0^\pi dv~{\cos v \over
[r^2 + z^2 +z_0^2 - 2 zz_0 \cos
v]^{1/2}}.
\label{eq:htwoonea}
\een
This one point function can be read off directly by considering the
leading contribution when $z = z_B \rightarrow 0$ :
\ben
<\cO>^{a^2=1}_{Poincare}  =  {z_B^2 z_0 \over 4(r^2 + z_0^2)^{3/2}}.
\label{eq:htwotwo}
\een
As expected, the one point function is a constant in
Poincar\'e time and the correct factor of cutoff appears.

\subsubsection{Supercritical Acceleration}

Finally we consider the case of $a^2 > 1$ given by the trajectory
(\ref{eq:trajone}) $\rho'  = \rho_0 $ in BTZ coordinates. Therefore we
first do the calculation in BTZ coordinates. The result is
(\ref{eq:htwofivea}) in Appendix II. In
Poincar\'e coordinates this becomes 
\ben
\Phi (X)  =  {z{\sqrt{\rho_0^2-1}}\over 2\pi}\int_0^\pi dv~
{\cos v \over [(\rho_0 t - z \cos v)^2 + (\rho_0^2
-1)(r^2+z^2-t^2)]^{1/2}},
\label{eq:htwosix}
\een
which leads to the one point function
\ben
<\cO>^{a^2>1}_{Poincare}  =  {z_B^2 \rho_0 {\sqrt{\rho_0^2-1}} \over 4}
{t \over [(\rho_0^2 -1)r^2 + t^2]^{3/2}}. 
\label{eq:htwoseven}
\een

\subsection{Global Boundary}

The one point functions in the boundary theory defined on the boundary
in global coordinates, $\nu = \nu_B \rightarrow 
{\pi \over 2}$ can be obtained by
entirely similar methods. For all cases we write down the field $\Phi
(X)$ in global coordinates and evaluate the leading nonzero
contribution in the limit $\nu_B \rightarrow {\pi \over 2}$. 

\subsubsection{Subcritical Accelerations}

For $a^2 < 1$ the calculation is trivial since the trajectories are
those of constant values of $\nu' = \nu_0$. The one point function
may be obtained by simply taking the limit of $\Phi (X)$ in
(\ref{eq:hten}) as $\nu
\rightarrow {\pi \over 2}$. The result is
\ben
<\cO>^{a^2<1}_{global}  =  {\cos^2 \nu_B \over 4}{\sin \nu_0 \over \cos^2
\nu_0}{\cos \phi \over [1 + \tan^2\nu_0 \sin^2 \phi]^{3/2}},
\label{eq:htwoeight}
\een
which is of course constant in global time. The overall factor of
$\cos^2 \nu_B$ is the UV cutoff factor as is evident from the form of
the metric and its appearance is consistent with the dimension of
the operator $\cO$.

\subsubsection{Critical Acceleration}

For $a^2 =1$, the field (\ref{eq:htwoone}) has to be first transformed
into global coordinates. The result is
\ben
\Phi (X) = {\cos \nu \over 2\pi}\int_0^\pi dv~{\cos v \over
\Delta^{1/2}},
\label{eq:htwonine}
\een
where
\ben
\Delta = (\sin^2\nu\sin^2\phi + \cos^2\nu) + z_0^2(\cos\tau + \sin\nu
\cos\phi)^2 - 2 z_0 \cos\nu \cos v~(\cos\tau + \sin\nu
\cos\phi).
\een
This leads to the one point function
\ben
<\cO>^{a^2=1}_{global}  =  {z_0\cos^2\nu_B \over 4}~{\cos \tau +
\cos \phi \over [\sin^2\phi + z_0^2 (\cos\tau + \cos\phi)^2]^{3/2}}.
\label{eq:hthirty}
\een
This is now a nontrivial function of the time $\tau$ on the global
boundary. 

\subsubsection{Supercritical Acceleration}

The field $\Phi (X)$ (\ref{eq:htwofivea}) is, in
terms of global coordinates,
\ben
\Phi (X) = {\cos\nu{\sqrt{\rho_0^2-1}}\over 2\pi}\int_0^\pi dv~
{\cos v \over [(\rho_0 \sin\tau - \cos\nu \cos v)^2 - (\rho_0^2
-1)(\sin^2\nu \cos^2\phi-\cos^2\tau)]^{1/2}},
\label{eq:hthreeone}
\een
which leads to the one point function
\ben
<\cO>^{a^2>1}_{global}  =  {\cos^2\nu_B~\rho_0 {\sqrt{\rho_0^2-1}} \over 4}
{\sin\tau \over [(\rho_0^2 -1)\sin^2\phi + \sin^2\tau]^{3/2}}.
\label{eq:hthreetwo}
\een

\subsection{Boundary in Hyperspherical coordinates}

In hyperspherical coordinates the
boundary is at $\mu = \mu_B$ and we have to consider the limit
$\mu_B \rightarrow \infty$.

\subsubsection{Subcritical Accelerations}

For $a^2 < 1$ we first need to write the field $\Phi (X)$ in
(\ref{eq:hseven}) at arbitrary
$X$ in the hyperspherical coordinates. The result is
\ben
\Phi (X) = {\sec \nu_0 \over 2\pi}\int_0^\pi dv{\cos v \over
\Sigma^{1/2}},
\label{eq:cfive}
\een
where
\ben
\Sigma = \sec^2\nu_0 (\sinh^2 \mu \sin^2 \theta \sinh^2\eta + \cosh^2 \mu) -
(\cos v + \sinh \mu \sin \theta \cosh \eta \tan \nu_0)^2 .
\een
The one point function is obtained by evaluating  the leading nonzero
contrinution as we take $\mu = \mu_B \rightarrow \infty$
\ben
<\cO>^{a^2 < 1}_{hyperspherical} = {1\over 4 \sinh^2 \mu_B}{\sin \nu_0
\over \cos^2 \nu_0} {\sin \theta \sinh \eta \over [\sec^2 \nu_0 +\sin^2 \theta
\sinh^2 \eta - \tan^2\nu_0 \sin^2 \theta]^{3/2}} .
\label{eq:csix}
\een

\subsubsection{Critical Acceleration}

The field $\Phi (X)$ in (\ref{eq:heighteen}) becomes in hyperspherical
coordinates 
\ben
\Phi (X) = {e^{-\eta}\over 2\pi} \int_0^\pi {\cos v \over
\tSigma^{1/2}},
\label{eq:cseven}
\een
where
\ben
\tSigma = e^{-2\eta}(1+\sinh^2 \mu \cos^2 \theta) + z_0^2 \sinh^2\mu
\sin^2 \theta - 2 z_0 e^{-\eta} \sinh \mu \sin \theta \cos v.
\een
The result for the one point function is
\ben
<\cO>^{a^2 = 1}_{hyperspherical} = {z_0 \over 4 \sinh^2 \mu_B}
{\sin \theta ~e^{-2\eta} \over [z_0^2 \sin^2\theta + e^{-2\eta} \cos^2
\theta]^{3/2}}. 
\label{eq:ceight}
\een

\subsubsection{Supercritical Acceleration}

For supercritical acceleration the trajectories with $\rho = \rho_0$
and $\psi = 0$ are in fact those with $\mu = \mu_0$ and $\theta = {\pi
\over 2}$ where $\cosh \mu_0 = \rho_0$. The field is given in 
(\ref{eq:htwofiveb}) which we rewrite below
\ben
\Phi (X) = {\sinh \mu_0 \over 2\pi} \int_0^\pi dv
{\cos v \over [\sinh^2 \mu \sinh^2 \mu_0 \sin^2 \theta - (\cosh \mu
\cosh \mu_0 - \cos v)^2]^{1/2}},
\label{eq:cnine}
\een
which leads to a one point function which is independent of the time
$\eta$, when expressed in terms of $\rho_0$ 
\ben
<\cO>^{a^2 > 1}_{hyperspherical}= {\rho_0 {\sqrt{\rho_0^2 -1}} \over 4
\sinh^2 \mu_B} { 1\over [\sin^2 \theta + \rho_0^2 \cos^2
\theta]^{3/2}}.
\label{eq:cten}
\een

\section{Boundary Interpretation of Unruh Radiation}

The one point functions calculated above provide us with holograms of
the accelerated trajectories. They have features expected from the
IR/UV correspondence. In each case, the one point function is
sharply peaked when the trajectory is close to the boundary. It is
clear from equations (\ref{eq:hthirteen}),(\ref{eq:htwotwo}) and
(\ref{eq:htwoseven}) that when the trajectories are given by $\nu_0 \sim
{\pi \over 2}, z_0 \sim 0$ and $\rho_0 \sim \infty$ respectively, the peak
appears at at $r=0$ for the Poincar\'e one point functions. Similarly
from (\ref{eq:twoeight}), (\ref{eq:thirty}) and (\ref{eq:threetwo}) it
is clear that the corresponding peaks appear at $\phi = 0$ for the
global one point functions. Finally, from (\ref{eq:csix}),
(\ref{eq:ceight}) and (\ref{eq:cten}) we see that there is a peak at
$\theta = {\pi \over 2}$. Recall that in the bulk the trajectories
all have $r=0$ or $\phi = 0$ or $\theta = {\pi \over 2}$. 
When the trajectories are at some finite
distance away from the boundary the one point functions have a spread
in the direction {\em along} the boundary and the spread increases as
the object in the bulk goes further away from the boundary.

How does a boundary field theorist figure out, by looking at these
profiles, whether a bulk observer along those trajectories will
perceive thermal radiation ?

We propose that the following is a possible way to do so. The idea is
to introduce a class of observers on the boundary according to whom
the one point function is time-independent. To do this it is crucial
to remember that the one point function is a $(1,1)$ operator in the
boundary theory. We thus go to a coordinate system in which the
transformed one point function is time independent.  These observers
are then co-moving with the profile of the trajectory on the
boundary. Generally, the time measured by these observers would be
different from the Poincar\'e, global or hyperspherical coordinate
times. Consequently the definition of particles would also be
different. We have to then figure out whether the vacuum on the
boundary appears as a vacuum defined by the positive frequency modes
with respect to this new time.  If not, there is a nontrivial
Bogoluibov transformation and one would generally perceive a mixed
state.

Making a coordinate transformation to implement this of course changes
the form of metric of the corresponding boundary theory. We will find
that the transformed form of the metric would be in general of the
form of cosmological spacetimes. The nature of this cosmology then
determines whether one will have a thermal state.

Note that the time in which these one point functions become
time-independent is {\em not} the comoving time of the resulting
cosmology. A further transformation to the latter comoving time would
reintroduce a time dependence in the one point function since the
latter is a $(1,1)$ tensor in the boundary theory.

\subsection{Poincar\'e Boundary}

First consider the description of the three classes of accelerated
trajectories in the bulk in the boundary theory defined by the
Poincar\'e boundary at $z = z_B \rightarrow 0$. We want to make coordinate
transformations on the boundary such that the one point functions 
(\ref{eq:hthirteen}),(\ref{eq:htwotwo}) and
(\ref{eq:htwoseven}) become time independent. The one point function 
(\ref{eq:htwotwo}) corresponding to critical acceleration is already
time independent. We have to thus deal with only the subcritical and
supercritical cases. To find these coordinate transformations we need
to use the fact that the operator $\cO$ is a $(1,1)$ operator in the
conformal field theory on the boundary. It is therefore best to use
null coordinates defined by
\ben
u = t-r,~~~~~~~~~\bu = t+r.
\label{eq:gone}
\een
We will also write the operator as $\cO_{u\bu}(u,\bu)$ in an obvious
notation.

\subsubsection{Subcritical Acceleration}

The following coordinate transformations to another set of null
coordinates $w, \bw$ renders the one point function
(\ref{eq:hthirteen}) constant
\ben
u = \tan {w \over 2},~~~~~~~~~~~~~\bu = \tan {\bw \over 2}.
\label{eq:gtwo}
\een
Using the transformation law
\ben
<\cO_{w\bw}>(w,\bw) = ({\partial u \over \partial w})({\partial \bu
\over \partial \bw}) <\cO_{u\bu}>(u,\bu),
\label{eq:gthree}
\een
and introducing
\ben
\btau = {1\over 2}(\bw + w)~~~~~~~~~\bphi = {1\over 2}(\bw - w),
\label{eq:gfour}
\een
one gets
\ben
<\cO_{w\bw}>^{a^2 < 1} = {z_B^2\over 4}{\sin \nu_0 \over \cos^2
\nu_0}{\cos \bphi \over [1 + \tan^2\nu_0 \sin^2 \bphi]^{3/2}},
\label{eq:gfive}
\een
which is independent of the new time $\btau$. The original metric of
the boundary
\ben
ds^2 = -dt^2 + dr^2,
\een
becomes
\ben
ds^2 = {1\over  (\cos \btau + \cos \bphi)^2}[- d\btau^2 + d\bphi^2]
\label{eq:gsix}
\een

This has the form of a cosmological spacetime with time dependent
metric. It is a fake cosmology, since the spacetime is really
flat. Nevertheless, in a way similar to the discussion of usual Unruh
radiation, we have to ask whether the modes which are positive
frequency with respect to $\btau$ are also purely positive frequency
with respect to $t$ and vice versa. In this case the answer is that
there is no mixing of positive and negative frequency modes. This is
because constant $\btau$ surfaces foliate the entire spacetime. The
vacuum which is defined with respect to positive frequency modes of
$t$ does not contain particles which are defined with respect to
positive frequency modes of $\btau$. The equivalence of Poincare and
global vacua have been shown in \cite{danielsson} from a slightly
different point of view.

In fact the transformed one point function (\ref{eq:gfive}) is exactly
identical to the one point function calculated in the global boundary
theory, equation (\ref{eq:htwoeight}) if me make the identifications
$\btau = \tau$ and $\bphi = \phi$ and if we identify the UV cutoff $z_B$
appearing in (\ref{eq:gfive}) with the UV cutoff $\cos \nu_B$ appearing
in (\ref{eq:htwoeight}). This is easy to understand from the bulk
point of view. While the value of the field $\Phi (X)$ produced by the
source is a scalar and independent of the coordinates, the one point
function extracted from it depends on the particular choice of the
boundary. The results for different choices should be related by the
conformal transformation which is the restriction of the bulk
coordinate transformation to the boundary. Indeed the coordinate
transformation (\ref{eq:gtwo}) is precisely that. Furthermore cutoffs
used in the two boundary theories should be related by comparing the
boundary metrics obtained from the bulk metric. This identifies $z_B$
with $\cos \nu_B$ for $z_B \rightarrow 0$ and $\nu_B \rightarrow {\pi \over
2}$.

Note that we could have made an overall scale transformation in addition
to (\ref{eq:gtwo}) and the one point function would remain time independent.
In the following we will show how to determine this overall scale,
which is important to extract the correct temeperature. We have not bothered
to do that in this case since there is no temperature anyway.

\subsubsection{Supercritical Acceleration}

In the bulk, an observer with supercritical acceleration detects
thermal radiation. To understand this from the point of view of the
theory on the Poincar\'e boundary, we need to find a coordinate
transformation which renders the one point function
(\ref{eq:htwoseven}) time independent. The appropriate coordinates are
now $(\baeta, \bpsi)$ defined by
\ben
r = {1\over \alpha}e^{-\alpha \baeta}\sinh \alpha \bpsi,
~~~~~~~~t = {1\over \alpha}e^{-\alpha\baeta}\cosh \alpha\bpsi,
\label{eq:gseven}
\een
where 
\ben
\alpha = {1\over {\sqrt{\rho_0^2-1}}},
\label{eq:gsevena}
\een
which is a conformal transformation.
The one point function now becomes
\ben
<\cO_{\baeta,\bpsi}>^{a^2>1}  =  {z_B^2 \rho_0  
\over 4 {\sqrt{\rho_0^2-1}}}
{\cosh \alpha \bpsi \over (\rho_0^2 \sinh^2 \alpha \bpsi + 1)^{3/2}},
\label{eq:geight}
\een
which is independent of the new time $\baeta$. In terms of these
coordinates, the metric on the Poincar\'e boundary becomes
\ben
ds^2 = {e^{-2\alpha\baeta}}[-d\baeta^2 + d\bpsi^2].
\label{eq:gnine}
\een
This is the metric of a two dimensional Milne universe. The latter is
given by the metric
\ben
ds^2 = e^{2b\zeta}(-d\zeta^2 + dx^2),
\label{eq:gten}
\een
and we have $b = -\alpha$ here. 
Of course we can make a transformation 
$\baeta \rightarrow \baeta$ to make $b = \alpha$.

As emphasized above the correct time for us is $\baeta$, which is the
conformal time of the Milne universe. This is {\em not}
the comoving time $\gamma = {1\over \alpha}e^{-\alpha\baeta}$ 
of the Milne universe. If we
make a further coordinate transformation to this time $\gamma$ the
resulting one point function, which is a tensor, will no longer be
indpendent of $\gamma$. 

The one point function (\ref{eq:geight}) is in fact identical to the
one point function obtained by considering the field $\Phi(X)$ and
going to the BTZ boundary at $\rho_B \rightarrow \infty$ and then
performing a scale transformation to identify
$\alpha \baeta = \eta$ and $\alpha \bpsi = \psi$. 
The cutoff $z_B$ has to be idetified with the cutoff
${1\over \rho_B}$. As expected, the coordinate transformation
(\ref{eq:gseven}) is the restriction of the coordinate transformation
relating BTZ and Poincar\'e coordinates to the Poincar\'e boundary, together
with an overall scale transformation.

The particular choice of $\alpha$ may be motivated in several ways. 
It is straightforward to see that for a given 
portion of the trajectory at some $\rho_0$, the elapsed proper time is
given by the expression
\ben
\lambda_{bulk} = {1\over z_B}\partial_{z_B}
\int dt dr <\cO>^{a^2 > 1}_{t,r}.
\label{eq:ngone}
\een
From the bulk point of view this follows from the equation satsified by the
field $\Phi (X,Y)$. However the expression (\ref{eq:ngone}) expresses this
entirely in terms of boundary quantities, viz. the 
derivative of the integral of the
one point function with respect to the cutoff. It may be easily checked that
the choice of $\alpha$ in (\ref{eq:gsevena}) then implies that
$\lambda_{bulk} = \baeta$.

Another way to see that (\ref{eq:gsevena}) is the correct choice is to 
see that after the conformal transformation to $(\baeta,\bpsi)$ the
integral of the one point function $<\cO>^{a^2>1}_{\baeta,\bpsi}$ over
space, viz. $\int d\psi <\cO>^{a^2>1}_{\baeta,\bpsi}$ is {\em independent of
the particular trajectory which is labelled by} $\rho_0$. Thus the overall
scaling we have used compensates for the redshift factor in the bulk.

It is well known that in the Milne universe, positive frequency modes
defined with respect to the time $\zeta$ mix with positive and
negative frequency modes defined with respect to the time $t$ and vice
versa \cite{birel}. Consider for example positive frequency modes of a
massive scalar field of mass $m$. The normalized positive frequency modes with
respect to the time $\zeta$ are 
\ben
\bchi_k = ({2b \over \pi} \sinh {\pi k \over b})^{-1/2}~J_{-{ik \over
b}}({m \over b}e^{b\zeta}),
\label{eq:geleven}
\een
while the positive frequency modes with respect to the time $t$ are
given by 
\ben
\chi_k = {1\over 2}({\pi \over b})^{1/2}~e^{\pi k \over 2b}~H^{(2)}_{ik
\over b}({m \over b}e^{b\zeta}),
\label{eq:gtwelve}
\een
where $J$ and $H^{(2)}$ denotes Bessel and Hankel functions
respectively. The fact that (\ref{eq:geleven}) are positive frequency
with respect to the time $\zeta$ is clear from the small $m$ expansion
of the Bessel function. To see why (\ref{eq:gtwelve}) are positive
frequency with respect to time $t$ consider a different time
coordinate $\gamma$ defined by
\ben
\gamma = {1\over b}e^{b\zeta}.
\een
A positive frequency mode with respect to the time $\gamma$ is
\ben
\phi_p = e^{-i(\omega_p \gamma - p x)}, ~~~~~~\omega_p^2 = p^2 + m^2.
\label{eq:gthirteen}
\een
The fourier transform of this in the $x$ space (the coordinate $x$ is
defined in (\ref{eq:gten})) is then easily seen to be, for $\gamma >
0$
\bea
\phi_k (\gamma) & = & \int dx e^{-ikx}~\phi_p \nn \\
& = & {1\over 2}({\pi \over b})^{1/2}~e^{\pi k \over 2b}~H^{(2)}_{ik
\over b}(m\gamma)~e^{-ik\alpha},
\label{eq:gfourteen}
\eea
where $\sinh \alpha b = {p \over m}$. This is exactly $\chi_k$ in
(\ref{eq:gtwelve}) upto a constant phase. The $m \rightarrow 0$ limit
is tricky : one has to always consider a finite $m$ and then take the
limit at the end.

The relationship between various Bessel functions can be now used to
show that 
\ben
\chi_k = {1\over 2 \sinh {\pi k \over b}}[e^{\pi k \over 2b} \bchi_k -
e^{-\pi k \over 2b} (\bchi_k)^*],
\label{eq:gfifteen}
\een
showing that the positive frequency mode with respect to the time $t$ 
is a linear combination of positive and negative frequency modes with
respect to the time $\zeta$. The Bogoliubov coefficients then imply
that the Poincar\'e vacuum (i.e. defined by modes $\chi_k$) is a thermal
state in terms of the $\zeta$-particles with a temperature
\ben
T = {b \over 2\pi}.
\label{eq:gsixteen}
\een
The underlying reason behind this is that the 
coordinates $\zeta,x$ cover only one wedge of the entire $(t,r)$
space, namely the wedge $t^2 > r^2$. The other two wedges are covered
by the well known Rindler coordinates. These can be obtained from the
Milne coordinates by performing the analytic continuation
\cite{tanaka} 
\ben
\chi_R = x - i{\pi \over 2},~~~~~~~\xi_R = i\gamma
\een
in the above metric. This leads to the standard Rindler metric with a
Rindler time $\chi_R$.

\begin{figure}
\centerline{
\epsfig{file=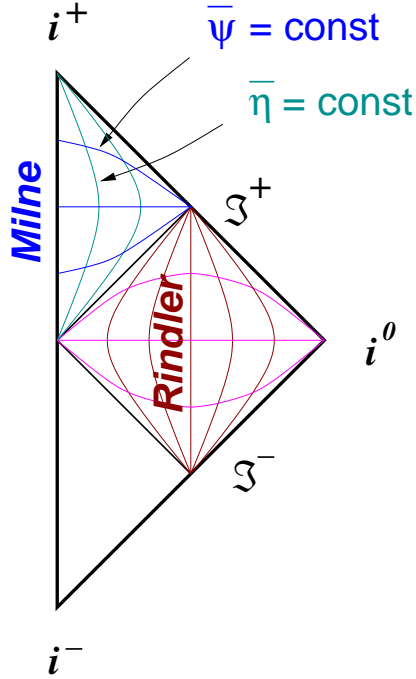,width=5.5 cm}}
\caption[Fig2]{The Penrose-Carter diagram of Poincar\'e boundary}
\end{figure}

Specializing to our case, we now see that the observers who use the time
$\baeta$ perceives a thermal state with a temperature
\ben
T = {1\over 2\pi {\sqrt{\rho_0^2-1}}},
\label{eq:ngtwo}
\een
which is the correct temperature detected by the accelerated observer in 
the bulk. 

From the point of view of the Poincar\'e boundary, the source moving on a
trajectory with supercritical acceleration is a pure state of the
conformal field theory, built upon the usual vacuum. However an
observer co-moving with the hologram of the source perceives this
vacuum as a mixed state and hence perceives all states constructed
over this vacuum as mixed states. Note that such observers are those
at constant values of the coordinate $\bpsi$ and are in fact described
in terms of the Poincar\'e coordinates $(r,t)$ by the motion
\ben
t = r \tanh \bpsi .
\een
This is motion with a constant velocity since the boundary metric is flat. 
The point, however, is that
the constant $\zeta$ surfaces are not Cauchy surfaces and one has to
also consider the spacelike surfaces of constant $\chi_R$ to perform a
quantization. The outgoing modes in the Rindler part of the spacetime
do not enter the Milne universe and this is the basic reason behind
thermal behavior.

The one point function (\ref{eq:htwoseven}) is nonzero in the region
which is not covered by the $\baeta, \bpsi$ coordinates. Observers who
use $\baeta$ as time cannot see the entire profile and therefore
performs an averaging, resulting in thermality.

Finally, the coordinate transformation (\ref{eq:gseven}) itself makes
it clear why in this case there is thermal behavior. This simply
arises from the fact that the transformation is periodic under 
$\baeta \rightarrow \baeta + 2\pi i{\sqrt{\rho_0^2-1}}$ 
and consequently all correlation
functions will be periodic with this imaginary period - a signature of
a thermal correlation function.

\subsection{Global Boundary}

From the point of view of the field theory defined on the global
boundary, we need to consider the critical and supercritical cases. 

\subsubsection{Critical Acceleration}

The coordinate transformation from $(\tau,\phi)$ to new coordinates
$(\tit,\tr)$ which makes the one point function (\ref{eq:hthirty})
independent of $\tit$ is the inverse of (\ref{eq:gtwo})
\ben
\tit - \tr = \tan {\tau - \phi \over 2},~~~~~\tit + \tr = \tan {\tau +
\phi \over 2},
\label{eq:gseventeen}
\een
and now the one point function becomes
\ben
<\cO_{\tit,\tr}> = {z_0 \cos^2\nu_B \over 4(\tr^2 + z_0^2)^{3/2}},
\label{eq:geighteen}
\een
which of course agrees with (\ref{eq:htwotwo}) after 
the identification $\tit = t$ and $\tr =
r$ and $\cos \nu_B = z_B$. The boundary metric now becomes
\ben
ds^2 = -d\tau^2 + d\phi^2 = {4  \over 4\tr^2 +
(1+\tit^2-\tr^2)^2}[-d\tit^2 + d\tr^2],
\label{eq:gnineteen}
\een
which is again a cosmological spacetime. However since the vacuum
defined in terms of $(t,r)$ is the same as that in terms of
$(\tau,\phi)$ (this is what we in fact showed in the previous
subsection), there is no thermal behavior.

\subsubsection{Supercritical Acceleration}

Proceeding as above it is clear that the coordinate system in which
the one point function becomes time independent is the BTZ system
restricted to the boundary. Thus we need to introduce the coordinates
$(\baeta, \bpsi)$ which are related to $\tau , \phi$ by
\ben
\tan \tau = {\cosh \alpha \bpsi \over \sinh \alpha \baeta},~~~~~~~~~~
\tan \phi = {\sinh \alpha \bpsi \over \cosh \alpha \baeta},
\label{eq:gtwenty}
\een
and the boundary metric now becomes
\ben
ds^2 = {2 \alpha^2
\over \cosh 2\alpha \baeta + \cosh 2\alpha \bpsi}[-d\baeta^2 + d\bpsi^2],
\label{eq:gtwoone}
\een
where $\alpha$ is given in (\ref{eq:gsevena}).
Once again the one point function becomes the same as that in BTZ
coordinates with the identification of the UV cutoffs $\rho_B \sim \sec
\nu_B$ and coordinates upto a scale 
$\alpha \baeta = \eta, \alpha \bpsi = \psi$. As discussed above the
spatial integral of the transformed one point function is now independent 
of $\rho=0$.

The metric (\ref{eq:gtwoone}) is again a cosmological
spacetime. Trajectories with constant $\psi$ are now {\em
accelerated} trajectories, but with a {\em non-uniform} acceleration
\ben
a_B^2 = {\cosh^2\psi \sinh^2\psi \over (\cosh^2 \eta + \sinh^2\psi)}.
\een
The acceleration vanishes in the far past and far future. Observers
moving along such a constant $\psi$ trajectory in fact perceive an
event horizon, pretty much like Rindler observers.

The origin of thermality is now a little different than the origin of
thermality in the Poincar\'e boundary theory. Here it is a bit similar,
though not identical to, the thermal behavior detected by Rindler
observers in flat space. Instead of the uniform acceleration of
Rindler observers, we now have a time dependent acceleration. Nevertheless
the physics is rather similar. Once again, the periodicity of the
transformation rules (\ref{eq:gtwenty}) is a signature of thermal
behavior with the correct temperature.

\subsection{Hyperspherical Boundary}

The boundary in hyperspherical coordinates is deSitter  spacetime. We
will consider the three classes of trajectories and follow the above
strategy to find the appropriate coordinates in which the one point
function becomes time independent. 

\subsubsection{Subcritical Acceleration}

The coordinate transformation from $\eta,\theta$ to new coordinates
$\ttau,\tphi$ is in fact that between hyperspherical and global
coordinates on the boundary
\ben
\tanh \eta = {\cos \ttau \over \cos \tphi},~~~~~~~\cos \theta = {\sin
\tphi \over \sin \ttau}.
\label{eq:celeven}
\een
With the identification $\ttau = \tau$ and $\tphi = \phi$ the one
point function becomes precisely (\ref{eq:htwoeight}) provided one
identifies the cutoffs as $\sinh \mu_B = \sec \nu_B$.

The boundary metric now becomes
\ben
ds^2 = {1 \over \sin^2 \ttau}[-d\ttau^2 + d\tphi^2].
\label{eq:ctwelve}
\een
The coordinates $\ttau, \tphi$ cover the entire 2d deSitter spacetime,
thus forming a global coordinate system. Boundary observers using this
time do not detect any thermal radiation in the invariant vacuum. 
This is what we expect from subcritical acceleration.

Once again the correct time $\ttau$ 
is not the comoving time $T = \cosh^{-1}(1/\sin \ttau)$. Observers
using this time $T$ {\em do} detect thermal radiation \cite{birel},
but this is not relevant to our discussion.

\subsubsection{Critical Acceleration}

For $a^2 =1$ one has to go to coordinates
\ben
e^{2\eta} = {1\over {\hat t}^2 - \hr^2},~~~~~~~~\cos \theta = {\hr \over
{\hat t}}.
\label{eq:cfifteen}
\een
As expected, the one point function becomes (\ref{eq:htwotwo}) with the
identification $\hr = r,~~{\hat t} = t$ and $\sinh \mu_B = {1\over
z_B}$. The deSitter  metric on the boundary now becomes
\ben
ds^2 = {1 \over {\hat t}^2}[-d{\hat t}^2 + d\hr^2].
\label{eq:csixteen}
\een
The coordinate $t = {\hat t}$ has the range $-\infty < t < \infty$ and 
of course cover the entire boundary (in fact
more). If the range of ${\hat t}$ is $-\infty < {\hat t} < 0$ rather than the
full range we would have the two dimensional steady state
universe \cite{birel}. What we have, instead, are two copies of steady state
universe which are attached to cover the entire $dS_2$ spacetime.
As such there is no thermal behavior.

\subsubsection{Supercritical Acceleration}

For supercritical acceleration, the one point function (\ref{eq:cten})
is already independent of the time $\eta$.
However, as explained above
we need to make a further scale transformation to coordinates $\teta,
\ttheta$ with $\eta = \teta /{\sqrt{\rho_0^2-1}}$ and
$\ttheta = \theta /{\sqrt{\rho_0^2-1}}$. Note that 
$\sin \mu_0 = {\sqrt{\rho_0^2-1}}$ so that this rescaling follows from
the form of the metric.
 
\begin{figure}
\centerline{
\epsfig{file=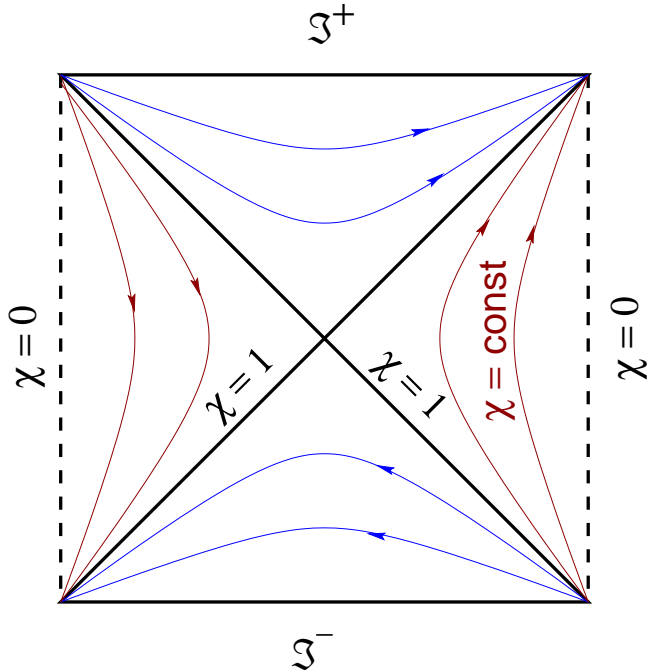,width=8.5 cm}}
\caption[Fig3]{The Penrose-Carter diagram of deSitter space}
\label{deSitter}
\end{figure}

The boundary metric may be
rendered into a more standard form by writing
$\cos {\theta \over {\sqrt{\rho_o^2-1}}} = \chi$ so that the metric becomes
\ben
ds^2 = \sin^2\mu_0[-(1-\chi^2) d\eta^2 + {d\chi^2 \over 1-\chi^2}].
\label{eq:ceighteen}
\een
Constant $\chi$ observers (see Fig.(\ref{deSitter}))
on the boundary thus have a horizon at $\chi
= 1$ and as a result they detect a thermal bath in the invariant
vacuum \cite{gibbhaw} with a temperature
( see also \cite{birel} and references to other
original work therein)
\ben
T = {1\over 2\pi\sin \mu_0} = {1\over 2\pi {\sqrt{\rho_o^2-1}}}.
\een
In particular, $\chi = 0$ is a geodesic - deSitter invariance then ensures that
all geodesic observers measure the same temperature.
The holographic interpretation of thermal
radiation detected by an object with supercritical acceleration in the
bulk is then interpreted as thermal radiation measured by stationary
observers in a deSitter spacetime on the boundary.

When the bulk object is moving along the holographic screen $\mu =
\mu_B$ itself, this is the fact used in \cite{msh} to obtain a
microscopic interpretation of two dimensional deSitter  entropy.

\section{Higher dimensions}

We have so far restricted our attention to $AdS_3$ for simplicity. The
discussion, however, can be easily generalized to any $AdS_{d+1}$. The
lesson we have learnt is that the boundary interpretation of Unruh
radiation in the bulk lies in the appearance of cosmological
spacetimes. From this point of view all we have to find is the
relevant cosmology. This is easily done for any $d$ since we have
also learnt that the class of observers who move with the hologram are
specified by a coordinate system which is in fact an appropriate
coordinate system in the bulk restricted to the boundary. 

The transformation rules to these coordinates are in fact {\em
identical} to those for $d=2$. This is because in our definitions
of the various coordinate systems in terms of the globally embedding
space of $(U,V,W_i)$ (equations (\ref{eq:three}),(\ref{eq:six}) and
(\ref{eq:nine})) we have used a common set of $(d-2)$ angles $\theta_1
\cdots \theta_{d-2}$. Using this fact we can immediately write down
the cosmology arising on the Poincar\'e boundary 
\bea
{\rm Subcritical}~~~ & ds^2 & = {1\over (\cos \btau + \cos
\bphi)^2}[-d\btau^2 + d\bphi^2 + \sin^2\bphi~d\Omega_{d-2}^2]. \nn \\
{\rm Supercritical}~~~ & ds^2 & = e^{-2\baeta}[-d\baeta^2 + d\bpsi^2 +
\sinh^2\bpsi ~d\Omega_{d-2}^2].
\label{eq:kone}
\eea
For the Global boundary we have the spacetimes
\bea
{\rm Critical}~~~& ds^2 & =  {4 \over 4\tr^2 +
(1+\tit^2-\tr^2)^2}[-d\tit^2 + d\tr^2 + \tr^2~d\Omega_{d-2}^2].\nn \\
{\rm Supercritical}~~~& ds^2 & = {2 \over \cosh 2\baeta + \cosh
2\bpsi}[- d\baeta^2 + d\bpsi^2 +
\sinh^2\bpsi ~d\Omega_{d-2}^2].
\label{eq:ktwo}
\eea

Finally for the $d$ dimensional deSitter boundary in hyperspherical
coordinates the following different metrics arise in the three
different cases
\bea
{\rm Subcritical}~~~ & ds^2 & = {1\over \sin^2 \ttau}[-d\ttau^2 +
d\tphi^2 
+ \sin^2\tphi~d\Omega_{d-2}^2]. \nn \\
{\rm Critical}~~~& ds^2 & =  {1\over {\hat t}^2}[-d{\hat t}^2 + d\hr^2 +
\hr^2~d\Omega_{d-2}^2]. \nn \\
{\rm Supercritical}~~~& ds^2 & = - \sin^2 \theta d\baeta^2 + d\theta^2 +
\cos^2 \theta ~d\Omega_{d-2}^2.
\label{eq:cnineteen}
\eea
In particular for the supercritical case one has $d$ dimensional Milne
universe on the Poincar\'e  boundary and the stationary form of the $d$
dimensional deSitter  metric on the hyperspherical boundary.

For $d\neq 2$ the one point functions can be also calculated using the
results of \cite{danielsson}.  However, the transformation properties
of the one point function would not be as simple as for $d=2$.

\section{Conclusions}

The main lesson of our study is that there is a dual relationship
between thermal behavior (or its absence) in cosmological spacetimes
on the boundary and thermal behavior observed by uniformly accelerated
objects in the bulk.

An interesting relationship between thermal effects due to
acceleration and those due to ``genunine'' gravity effects has been
uncovered in \cite{deser}. Here one
considers some {\em flat} global embedding spacetime, typically with
more than one timelike dimension, in which various nontrivial lower
dimensional spacetimes, like deSitter , anti-deSitter  or various
kinds of black holes, are embedded by defining a suitable
hypersurface. Accelerated trajectories in these lower dimensional
spacetimes - e.g. those in $AdS$ spacetimes discussed in this paper,
or those of stationary observers in a deSitter  space or black hole
background - then become uniformly accelerated trajectories in the
higher dimensional flat emebdding spacetime. Thermal effects of the
former, which may be thought of as ``genuine'' gravity effects, can be
then obtained from Unruh effect in the embedding spacetime. {\em Note
that in this discussion the trajectories in the lower dimensional
nontrivial spacetime are not holograms of those in the embedding space
- they are the latter trajectories themselves.} In other words the
trajectories always lie in the lower dimensional space.

If we consider the $AdS$ space time of this paper as the global
embedding spacetime, and the various boundaries are used to define
lower dimensional spacetime, the situation may appear somewhat
similar to the setup of \cite{deser}. However, the problem we have
studied is rather different. In our work the trajectories in fact {\em
do not lie on the boundary}. Rather they are transverse to the
boundary and their holograms are
expanding blobs on the boundary. The thermal behavior in the
bulk now has an interpretation in terms of observers in appropriate
cosmologies. 

In one sense the problem addressed in \cite{msh} also appears similar
to that in \cite{deser}. Here one considers the induced deSitter space
on a constant $\mu$ surface in hyperspherical coordinates in
$AdS_3$. Stationary observers in this deSitter space then become
uniformly accelerated observers in the $AdS$ space with supercritical
acceleration.  However the crucial difference with \cite{deser} 
is that this constant
$\mu$ surface, {\em on which the bulk trajectory lies}, is regarded
as a holographic screen. Furthermore as opposed to the usual AdS/CFT
setup (which is the one used in our work), $\mu$ is finite and {\em
dynamically determined} in terms of the tension of a 1-brane which is
on this surface. For the same reason, the holographic theory has
dynamical gravity
\cite{randall}. This allows a {\em microscopic} understanding of the
thermodynamics measured by these stationary observers.

In our work we have argued that a similar correspondence between Unruh
thermal properties of bulk objects {\em moving transverse to the boundary}
and cosmological thermal properties on the boundary theory. The
boundary is of course not the trajectory itself.  Nevertheless, the
presence or absence of thermal behavior follows from the appropriate
choice of time in the boundary cosmology.  This correspondence is in
the traditional AdS/CFT framework and different choices of the
boundary theory give rise to different cosmologies, which of course
include deSitter  spacetimes.

Finally we note that we have not obtained these conclusions from the
boundary theory itself. Rather we have simply presented what the
AdS/CFT predicts in the boundary theory. The outstanding task is to
obtain these results starting from the specific CFT on the
boundary. That would lead to a microscopic understanding. As usual,
the problem is that one has to compute quantities at strong
coupling. Nevertheless one may hope that some qualitative features can
be nevertheless derived.

\section{Acknowledgements} We would like to thank 
the High Energy Theory group and Cosmology and General Relativity
group at YITP for providing a stimulating atmosphere and T. Yoneya for
a discussion.  A.Z. is also grateful to the Killam Trust for its
support. S.R.D would like to thank the Theory group of University of
Hokkaido for hospitality during the final stages of preparation of
this manuscript.

\section{Appendix I : Coordinate systems in $AdS_{d+1}$}

All the three coordinate systems we have used in this paper 
are various ways of solving the basic
definition of $AdS_{d+1}$ as a hyperboloid in $d+2$ dimensional flat
space with two timelike coordinates. The coordinates in this embedding
spacetime are denoted by $U,V, W_i$ with $i = 1, \cdots D$. The metric
in this emebdding space is
\ben
ds^2 = -dU^2-dV^2 +\sum_{i=1}^d dW_i^2
\label{eq:one}
\een
The $AdS_{d+1}$ is defined as the hyperboloid
\ben
-U^2-V^2 +\sum_{i=1}^d W_i^2 = -R^2
\label{eq:two}
\een
In the following we will measure all distances in units of $R$, so we
will set $R=1$.

\subsection{Global Coordinates}

A global coordinate system on $AdS_{d+1}$ is defined as 
\bea
U & = & \sec \nu~\cos \tau \nn \\
V & = & \sec\nu~\sin \tau \nn \\
W_1 & = & \tan \nu~\cos \phi \nn \\
W_2 & = & \tan \nu~\sin \phi~\cos \theta_1 \nn \\
\cdots \nn \\
W_d & = & \tan \nu~\sin\phi \sin \theta_1 \cdots \sin \theta_{d-2}
\label{eq:three}
\eea
The ranges of the coordinates are
\bea
0 & < & \nu < \pi/2 \nn \\
0 & < & \tau < 2\pi \nn \\
0 & < & \phi, \theta_i < \pi ~~~~~~~~(i=2,\cdots (d-3))\nn \\
0 & < & \theta_{d-2} < 2\pi 
\label{eq:four}
\eea

These coordinates cover the entire $AdS_{d+1}$. It is sometimes 
conveninent to consider the covering space where
$\tau$ ranges from $-\infty$ to $\infty$ by attaching identical
copies. The metric follows from substituting (\ref{eq:three}) in
(\ref{eq:one}) 
\ben
ds^2 = \sec^2\nu ( - d\tau^2 + d\nu^2 + \sin^2\nu d\Omega_{d-1}^2)
\label{eq:afive}
\een
Surfaces of constant $\nu$ have spatial sections which are $S^{d-1}$.
The global boundary is at $\nu = \nu_B \rightarrow \pi/2$ with the metric
\ben
ds_{gB}^2 = -d\tau^2 + d\Omega_{d-1}^2
\een

\subsection{Poincar\'e Coordinates}

The Poincar\'e coordinate system is defined by the relationships
\bea
U & = & {1\over 2z}[1+(r^2+z^2-t^2)] \nn \\
V & = & {t \over z}\nn \\
W_1 & = & {1\over 2z}[1-(r^2+z^2-t^2)] \nn \\
W_2 & = & {r\over z} \cos \theta_1 \nn \\
\cdots \nn \\
W_3 & = & {r\over z} \sin\theta_1 \cos\theta_2 \nn \\
W_d & = & {r\over z}\sin\theta_1 \sin\theta_2 \cdots \sin \theta_{d-2}
\label{eq:six}
\eea
where the ranges of the coordinates are
\bea
-\infty & < & t < \infty \nn \\
0 & < & z < \infty \nn \\
0 & < & r < \infty \nn \\
0 & < & \theta_i < \pi ~~~~~~~~(i=2,\cdots (d-3))\nn \\
0 & < & \theta_{d-2} < 2\pi 
\label{eq:seven}
\eea

The metric is now
\ben
ds^2 = {1\over z^2}[-dt^2 + dz^2 + dr^2 + r^2 d\Omega^2_{d-2}]
\label{eq:aeight}
\een
The Poincar\'e system does not cover the entire $AdS$ spacetime. There is a
horizon at $z=\infty$. 

Surfaces of constant $z$ are $d$ dimensional flat spacetimes.
The Poincar\'e boundary is at $z=z_B \rightarrow 0$ and has the
standard flat metric upto an overall constant 
\ben
ds_{PB}^2 = -dt^2 + dr^2 + r^2 d\Omega^2_{d-2}
\label{eq:aeighta}
\een

\subsection{``BTZ'' Coordinates}

The third coordinate system we will use will be referred to ``BTZ''-type
coordinates since this is a natural extension of the BTZ coordinates for
$d=2$. These are defined by
\bea
U & = & {\sqrt{\rho^2-1}}\sinh \eta \nn \\
V & = & \rho \cosh \psi \nn \\
W_1 & = & {\sqrt{\rho^2-1}}\cosh \eta  \nn \\
W_2 & = & \rho \sinh\psi \cos \theta_1 \nn \\
\cdots \nn \\
W_3 & = & \rho \sinh\psi \sin\theta_1 \cos\theta_2 \nn \\
W_d & = & \rho \sinh\psi\sin\theta_1 \sin\theta_2 \cdots \sin \theta_{d-2}
\label{eq:nine}
\eea
where the ranges of the coordinates are
\bea
-\infty & < & \eta < \infty \nn \\
0 & < & \psi < \infty \nn \\
1 & < & \rho < \infty \nn \\
0 & < & \theta_i < \pi ~~~~~~~~(i=2,\cdots (d-3))\nn \\
0 & < & \theta_{d-2} < 2\pi 
\label{eq:ten}
\eea
The metric is now
\ben
ds^2 = -(\rho^2-1)d\eta^2 + {d\rho^2 \over (\rho^2-1)} + \rho^2[d\psi^2  +
\sinh^2\psi d\Omega_{d-2}^2]
\label{eq:aeleven}
\een
These coordinates also do not cover the entire $AdS$ spacetime and has
horizons at $\rho^2 = 1$.

Surfaces of constant $\rho$ have spatial sections which are
hyperbolic spaces in $(d-1)$ dimensions. 
The boundary is at $\rho = \rho_B \rightarrow 
\infty$ and has the metric
\ben
ds^2_{BTZB} = -d\eta^2 + d\psi^2  +
\sinh^2\psi d\Omega_{d-2}^2
\label{eq:atwelve}
\een

\subsection{Hyperspherical Coordinates}

These coordinates are analytic continuations of the standard polar
coordinates on a $(d+1)$ dimensional sphere. In terms of the embedding
space one has
\bea
U & = & \sinh \mu \sin \theta \sinh \eta \nn \\
V & = & \cosh \mu \nn \\
W_1 & = & \sinh \mu \sin \theta \cosh \eta \nn \\
W_2 & = & \sinh \mu \cos \theta \cos \theta_1 \nn \\
\cdots \nn \\
W_3 & = & \sinh \mu \cos \theta \sin\theta_1 \cos\theta_2 \nn \\
W_d & = & \sinh \mu \cos \theta 
\sin\theta_1 \sin\theta_2 \cdots \sin \theta_{d-2}
\label{eq:ctwenty}
\eea
The time is the same as in BTZ coordinates and so are the angles
$\theta_i$ with $i = 1,\cdots (d-2)$.
The ranges of the other two coordinates are
\bea
0 & < & \mu < \infty \nn \\
1 & < & \theta < \pi \nn \\
\label{eq:ctwoone}
\eea
The metric is now
\ben
ds^2 = -\sinh^2 \mu \sin^2 \theta d\eta^2 + d\mu^2
+ \sinh^2 \mu [~d\theta^2  +
\cos^2\theta~d\Omega_{d-2}^2~] 
\label{eq:caeleven}
\een
These coordinates do not cover the entire $AdS$ spacetime. Observers
at a given point in space have a horizon at $\theta = {\pi \over 2}$. 

Surfaces of constant $\mu$ are $d$ dimensional deSitter  spacetimes.
The boundary is at $\mu = \mu_B \rightarrow \infty$ and has a
stationary metric
\ben
ds^2_{HypB} = - \sin^2 \theta d\eta^2 + d\theta^2 +
\cos^2\theta~d\Omega_{d-2}^2
\label{eq:catwelve}
\een

\subsection{Coordinate transformations}

The transformation rules between the various coordinates may be easily
obtained by comparing their definitions in terms of the basic
embedding coordinates $U,V,W_i$. Since we have chosen a common set of
$(d-2)$ angles $\theta_i$ in all the three coordinate systems, this
job is particularly simple : one has to just relate the sets
$(\tau,\nu,\phi)$, $(t,z,r)$ ,$(\eta,\rho,\psi)$ and
$(\eta,\mu,\theta)$.  As a result these rules are same for all values
of $d$. These formulae can be easily worked out from the above
expressions.

\section{Appendix II : Calculation of the field $\Phi (X)$}

\subsection{Subcritical Accelerations, $a^2 < 1$}

First consider trajectories given by (\ref{eq:trajthree}) which have
accelerations $a^2 < 1$. 
Since these correspond to constant values of
$\nu = \nu_0$ it is most conveninent to write down the Green's
function and the field $\Phi (X)$ in global coordinates. In
(\ref{eq:hfive}) we thus have
\ben
Y(\lambda) : \nu ' = \nu_0~~~\phi' = 0
\label{eq:hseven}
\een
The proper time interval $d\lambda$ along the trajectory is given by
\ben
d\lambda = \sec \nu_0 d\tau' = dv~\sec \nu_0~{d\tau' \over dv}
\label{eq:height}
\een
Using (\ref{eq:htwo}) we get
\ben
{d\tau' \over dv} = {\sin v \over 
[\sec^2 \nu \sec^2 \nu_0 - (\cos v + \tan
\nu \tan \nu_0 \cos \phi)^2]^{1/2}}
\label{eq:hnine}
\een
Using (\ref{eq:hone}) one gets
\ben
\Phi (X) = {\sec\nu_0 \over 2\pi} 
\int_0^\pi dv~{\cos v \over [\sec^2 \nu \sec^2 \nu_0 - (\cos v + \tan
\nu \tan \nu_0 \cos \phi)^2]^{1/2}} 
\label{eq:hten}
\een

\subsection{Critical Acceleration}

For the trajectory (\ref{eq:trajtwo}) with critical acceleration one
has
\ben
Y(\lambda) : z'=z_0~~~r=0
\label{eq:heighteen}
\een
The proper time interval is given by
\ben
d\lambda = {1\over z_0}dt' = dv~{1\over z_0}~{dt' \over dv}
\label{eq:hnineteen}
\een
and from (\ref{eq:hthree}) we obtain
\ben
{dt' \over dv} = {z z_0\sin v \over [r^2 + z^2 +z_0^2 - 2 zz_0 \cos
v]^{1/2}}
\label{eq:htwenty}
\een
The field at a point $X = (t,r)$ then becomes
\ben
\Phi (X)  =  {z \over 2\pi}\int_0^\pi dv~{\cos v \over
[r^2 + z^2 +z_0^2 - 2 zz_0 \cos
v]^{1/2}}
\label{eq:htwoone}
\een

\subsection{Supercritical Acceleration}

A $a^2 > 1$ trajectory
(\ref{eq:trajone}) is best described in BTZ coordinates as
\ben
Y (\lambda) : \rho' = \rho_0~~~~\psi = 0
\label{eq:htwothree}
\een
or equivalently in hyperspherical coordinates as
\ben
Y (\lambda) : \mu = \mu_0~~~~~~\theta = {\pi \over 2}
\label{eq:htwothreea}
\een
with
\ben
\rho_0 = \cosh \mu_0
\label{eq:htwothreeb}
\een
The proper time is now given by
\ben
d\lambda = {\sqrt{\rho_0^2 -1}}~d\eta' = {\sqrt{\rho_0^2 -1}}~dv~{d\eta'
\over dv}
\label{eq:htwofour}
\een
From the expression (\ref{eq:hfour}) one has
\bea
{d\eta' \over dv} &=&{\sin v \over [(\rho_0 V - \cos v)^2 +
(\rho_0^2 -1)(U^2-W_1^2)]^{1/2}}  \nn\\
&=& {\sin v \over [(\rho\rho_0 \cosh \psi - \cos v)^2 -
(\rho^2 -1 )(\rho_0^2 -1)]^{1/2}}
\label{eq:htwofive}
\eea
Using the formula for the geodesic distance in BTZ coordinates we have
\ben
\Phi (X) = {{\sqrt{\rho_0^2 - 1}}\over 2\pi}\int_0^\pi
dv {\cos v \over [(\rho\rho_0 \cosh \psi - \cos v)^2 -
(\rho^2 -1 )(\rho_0^2 -1)]^{1/2}}
\label{eq:htwofivea}
\een
Similarly in hyperspherical coordinates we have 
\ben
\Phi (X) = {\sinh \mu_0 \over 2\pi} \int_0^\pi dv
{\cos v \over [\sinh^2 \mu \sinh^2 \mu_0 \sin^2 \theta - (\cosh \mu
\cosh \mu_0 - \cos v)^2]^{1/2}}
\label{eq:htwofiveb}
\een

\subsection{Transformation to appropriate coordinates}

To extract the one point function for a particular boundary we need
the expressions for $\Phi (X)$ in appropriate coordinates. As
emphasized above, in the bulk $\Phi (X)$ is a scalar, so all we need
to do is to rewrite the expressions above in these coordinates. This
is easily done by using the definitions in Appendix I in terms of the
embedding coordinates $(U,V,W_i)$.

\end{document}